\documentclass[referee]{aa}

\usepackage{graphicx}
\usepackage{natbib}
\usepackage{latexsym}
\usepackage{amsmath}
\usepackage{amssymb}
\usepackage{amstext}
\usepackage{color}
\usepackage{psfrag}

\def\k{km s$^{-1}$}
\def\ks{km s$^{-1}$~}

\def\m{$^\prime$}
\def\s{$^{\prime\prime}$}

\def\cm3{cm$^{-3}$}

\def\2{$^{12}$CO}
\def\3{$^{13}$CO}

\def\msol{M$_\odot$}

\begin{document}

\title{The environs of the ultracompact HII region G45.45+0.06} 
\author {S. Paron \inst{1}
\and S. Cichowolski \inst{1}
\and M. E. Ortega  \inst{1}
}

\institute{Instituto de Astronom\'{\i}a y F\'{\i}sica del Espacio (IAFE),
             CC 67, Suc. 28, 1428 Buenos Aires, Argentina\\
             \email{sparon@iafe.uba.ar} 
}
\offprints{S. Paron}

   \date{Received <date>; Accepted <date>}

\abstract{}{G45.45+0.06 is an ultra-compact HII (UCHII) region extensively studied. 
It is known that G45.45+0.06 is embedded in a complex of UCHII regions, but up the date, the 
surrounding ISM in a larger spatial scale has not been analyzed.}
{Using data from large-scale surveys: 
Two Micron All Sky Survey, GLIMPSE, MIPSGAL, MAGPIS and GRS, we performed a multiwavelength study of a region
about 7\m $\times$ 7\m~in the vicinity of G45.45+0.06.}
{We found that the UCHII complex lies in a border of a larger (diameter of $\sim$ 3\m) and
fainter HII region, which is located at the same distance as G45.45+0.06, $\sim 8$ kpc. In this work, this 
larger HII region was called G45L. 
A good morphological correlation is observed between the PDRs 
and the molecular gas mapped in the \3 J=1--0 and CS J=2--1 lines, suggesting that G45L may be collecting the molecular 
material.
From a near- and mid-IR photometric study, we found three sources, likely O-type stars, that are possibly 
responsible for the creation of G45L. Additionally we found several YSO candidates lying 
preferently in the molecular shell that surrounds G45L. 
Our results confirm that the region southeastern the UCHII complex where G45.45+0.06 is embedded 
and eastern G45L is active in star formation.
We suggest that G45L has been expanding during about $2 \times 10^6$ yr and 
could have triggered the formation of the zero-age main sequence stars that are ionizing the UCHII 
region G45.45+0.06. However we can not discard that both HII regions are coeval. 

}{}

\titlerunning{The environs of the UCHII region G45.45+0.06}
\authorrunning{S. Paron et al.}


\maketitle

\section{Introduction}

Nowadays it is well established that the formation of massive stars can be
triggered by the action of expanding HII regions through the ``collect and collapse'' process.
During its supersonic expansion, an HII region can collect a dense layer
of material between its ionization and shock fronts. This layer can be
fragmented in massive condensations that then may collapse to lead to the
formation of new stars. Thus, it is expectable the presence of protostars,
young stars, and ultra-compact HII (UCHII) regions  on the borders of HII
regions.
Several observational evidence have been found supporting this star
forming mechanism (see e.g. \citealt{poma09,zav07}, and reference therein).

G45.45+0.06 is a luminous Galactic UCHII region that has been extensively studied in the radio 
continuum and in molecular lines (e.g. \citealt{wood89,garay93,wilner96,hatchell98}). This UCHII 
region presents CH$_{3}$OH and H$_{2}$O maser emission \citep{codella04}. Recently, G45.45+0.06 was included in 
the Boston University Catalog of Galactic HII Region Properties \citep{anderson09}. Using the \3 J=1--0 
emission obtained from the GRS\footnote{Galactic Ring Survey \citep{jackson06}}, the authors derive a 
v$_{\rm LSR} \sim 55.6$ \ks and 
a kinematical distance of 8.2 kpc for this UCHII region, in agreement, within errors, with previous estimations (e.g. 
v$_{\rm LSR} \sim 55.9$ \ks and d $\sim$ 7.7 kpc, \citealt{kolpak03}). 
As early proposed by \citet{matt77}, G45.45+0.06 is part 
of a cluster of several UCHII regions. \citet{giveon05,giveon05b} generated a 
catalog matching VLA Galactic plane catalogs at 5 and 1.4 GHz with new radio continuum observations and the MSX6C 
Galactic plane catalog. According to this catalog, G45.45+0.06 is part of a complex of five radio compact HII regions.
\citet{feldt98} performed a near- and mid-infrared study of G45.45+0.06 and concluded that this UCHII region, the
oldest in the complex, is a young OB cluster around which sequential star formation is taking place. Recently,
\citet{blum08} using NIFS behind ALTAIR on Gemini North identified several massive O-type stars that are ionizing
G45.45+0.06.
The complexity of this region is evident. 
Moreover, in the vicinity of this complex (see Figure \ref{present}),
\citet{cyga08} discovered an ``extended green object'' (EGO), a source with extended
{\it Spitzer}-IRAC 4.5 $\mu$m emission which is usually presented in green colour. According to the authors, 
an EGO is a massive young stellar object (MYSO) driven outflows. The presence of this EGO reveals that  
star formation is taking place in this region. Indeed, this EGO coincides with the high-mass star forming region
G45.47+0.05 studied by \citet{remijan04}.

In this work, we present a multiwavelength study of the spatial environment surrounding 
the UCHII region G45.45+0.06, with the purpose of exploring the ISM around the complex. We use 
survey and archival data to show that G45.45+0.06 is located in the environs of a larger HII region
whose expansion could have originated the conditions for the formation of the UCHII region G45.45+0.06.

\section{Data}

The data presented and analyzed in this work were extracted from five large-scale surveys: 
Two Micron All Sky Survey (2MASS)\footnote{2MASS is a joint project of
the University of Massachusetts and the Infrared Processing and Analysis Center/California Institute of Technology,
funded by the National Aeronautics and Space Administration and the National Science Foundation.}, 
Galactic Legacy Infrared Mid-Plane Survey Extraordinaire (GLIMPSE), 
MIPSGAL, MAGPIS and GRS. GLIMPSE is a mid infrared survey of the inner Galaxy performed using the {\it Spitzer 
Space Telescope}. We used the mosaiced images from 
GLIMPSE and the GLIMPSE Point-Source Catalog (GPSC) in the {\it Spitzer}-IRAC (3.6, 4.5, 5.8 and 8 $\mu$m). 
IRAC has an angular resolution between 1\farcs5 and 1\farcs9 (see \citealt{fazio04} and \citealt{werner04}).
MIPSGAL is a survey of the same region as GLIMPSE, using MIPS instrument (24 and 70 $\mu$m) on {\it Spitzer}. 
The MIPSGAL resolution at 24 $\mu$m is 6\s. MAGPIS is a radio continuum survey of the 
Galactic plane at 6 and 20 cm using the VLA in configurations B, C and D combined with the Effelsberg 100 m 
single-dish telescope \citep{white05}. The GRS is being performed by the Boston University and the 
Five College Radio Astronomy Observatory (FCRAO). The survey maps the Galactic Ring in the \3 J=1--0 line
with an angular and spectral resolution of 46\s~and 0.2 \k, respectively (see \citealt{jackson06}). 
The observations were performed in both position-switching and On-The-Fly mapping modes, achieving an 
angular sampling of 22\s. We also analyzed
the additional data that this survey presents: the CS J=2--1 line, which has similar angular and spectral resolutions
as the \3 J=1--0 line. 

\section{Results and discussion}

\begin{figure}[h]
\centering
\includegraphics[width=8cm]{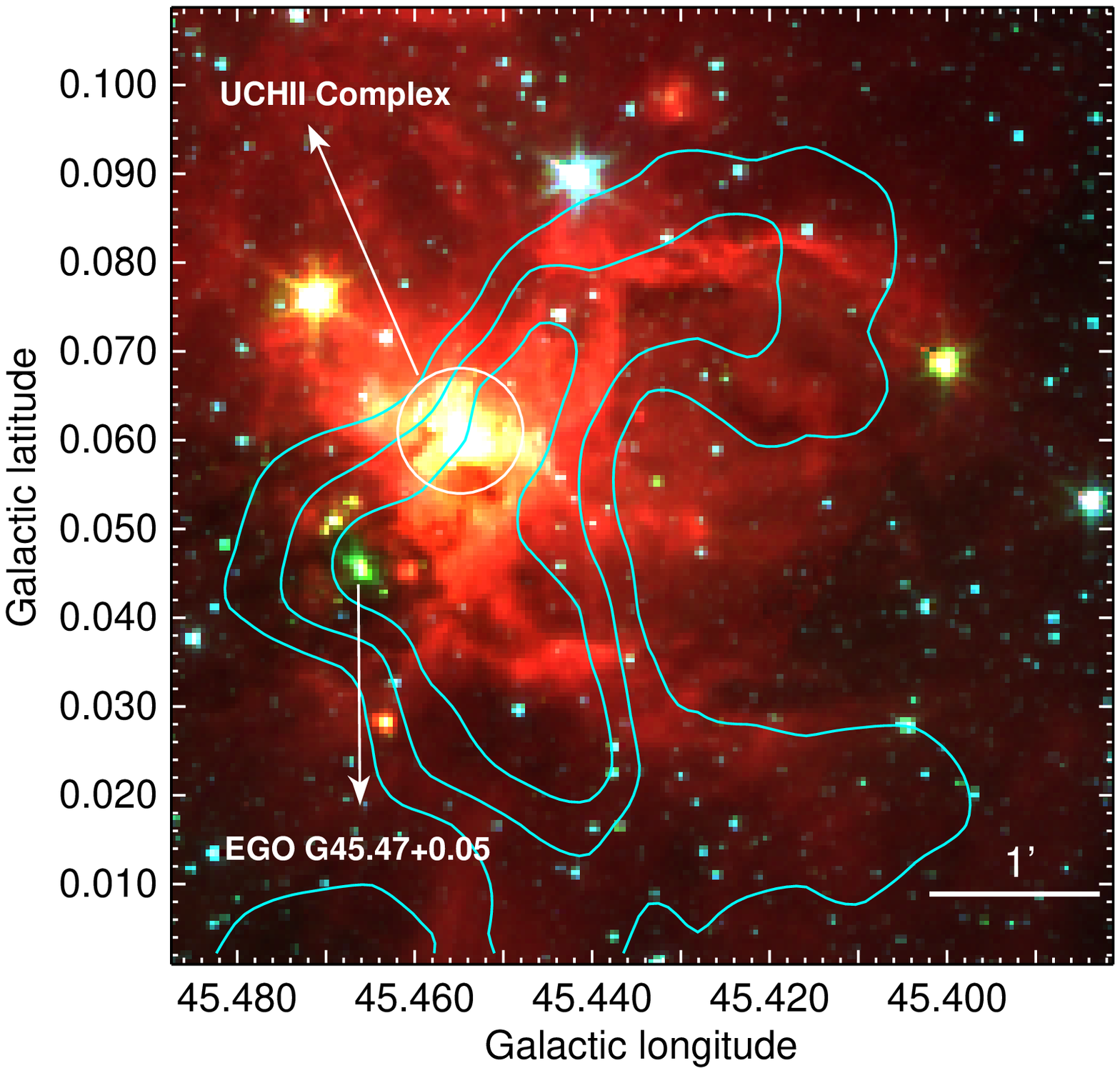}
\caption{{\it Spitzer}-IRAC three color image (3.5 $\mu$m $=$ blue, 4.5 $\mu$m $=$ green and 8 $\mu$m $=$ red). The
contours correspond to the \3 J=1--0 emission as extracted from the GRS integrated between 50 and 65 \k, the levels
are 27, 37 and 48 K \k. The UCHII complex is remarked with a circle. Also, the EGO discovered by \citet{cyga08}
is indicated. The angular resolutions are $\sim$ 1\farcs5 and $\sim$ 46\s~for the {\it Spitzer} and molecular data,
respectively.}
\label{present}
\end{figure}

Figure \ref{present} shows a {\it Spitzer}-IRAC three color image as extracted from GLIMPSE of a region
about 7\m $\times$ 7\m~in the vicinity of the UCHII complex where G45.45+0.06 lies.
The three IR bands are 3.6 $\mu$m (in blue), 4.5 $\mu$m (in green) and 8 $\mu$m (in red). The contours correspond
to the \3 J=1--0 emission as extracted from the GRS integrated between 50 and 65 \k, a velocity range around the
v$_{LSR }$ of G45.45+0.06. The UCHII complex lies on a border of a more extended structure as
seen mainly in the {\it Spitzer}-IRAC 8 $\mu$m band.
The morphological correlation between the 8 $\mu$m, mainly originated in the policyclic aromatic hydrocarbons 
(PAHs), and the molecular gas is suggestive of an expansive and ``collective'' HII region. It is important 
to note that the PAHs emission 
delineates the HII region boundaries. This is because
these molecules are destroyed inside the ionized region, but are excited in the photodissociation region (PDR) by 
the radiation leaking from the HII region \citep{poma09}.  

\begin{figure*}[tt]
\centering
\includegraphics[width=14cm]{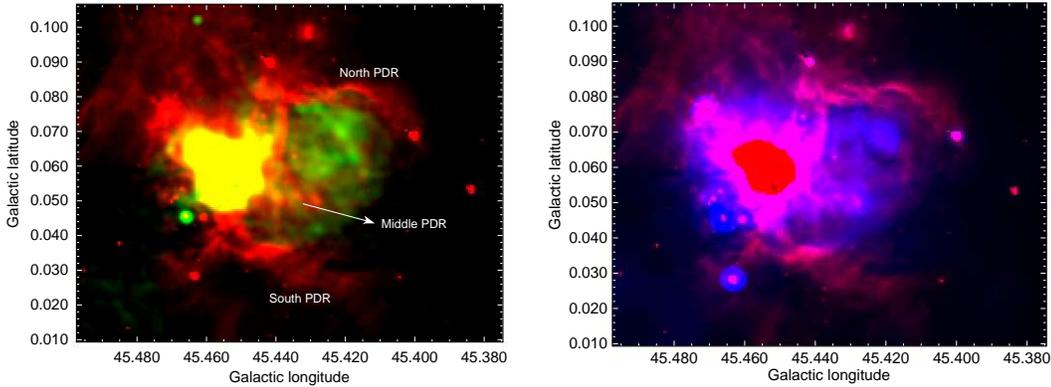}
\caption{Two color images. Left: the red is the {\it Spitzer}-IRAC 8 $\mu$m emission and the green is the above
3$\sigma$ radio continuum emission at 20 cm extracted from MAGPIS. Yellow is the superposition of the IR
and radio continuum emissions. 
Right: the red is the same as the left image and the blue represents the MIPSGAL emission at 24 $\mu$m. The red
structure onto the UCHII complex is not real, it is just due to the presence of saturated pixels of the MIPSGAL
emission.}
\label{combi}
\end{figure*}

Although G45L is not completely bordered by a PDR, its morphology resembles the structure of
the IR dust bubbles associated with O and early-B type stars: a PDR visible in
the 8 $\mu$m band, which encloses ionized gas observed at 20 cm and hot dust observed at 24 $\mu$m
(see \citealt{church06,church07,watson08}). 
Figure \ref{combi} (left) displays a composite two color image, where the red represents the {\it Spitzer}-IRAC
8 $\mu$m band and the green the above 3$\sigma$ radio continuum emission at 20 cm extracted from MAGPIS. 
The high angular resolution of the MAGPIS 20 cm image allows us to distinguish two different radio continuum structures: 
that associated with the UCHII 
complex, where both colors are combined and satured (yellow), and other more diffuse towards the West, enclosed by the 
8 $\mu$m emission that indicates the photodissociation regions (PDRs). Hereafter, this latter structure, 
an HII region of about 3\m~in diameter, 
possible related to the molecular gas shown in Figure \ref{present} as contours, will be called G45L.
As indicated in Figure \ref{combi} (left), the brightest part of the radio continuum diffuse structure located 
at $l \sim 45$\fdg$42, b \sim 0$\fdg$072$ is bordered by two PDRs: the North and the middle PDRs. Towards the 
South, the HII region is fainter and bordered by what we called South PDR.
Figure \ref{combi} (right) shows a two color image where 
again the red is the {\it Spitzer}-IRAC 8 $\mu$m emission and the blue represents the MIPSGAL emission at 24 $\mu$m. 
The red structure observed onto the UCHII complex is not real, it is just due to the presence of saturated pixels 
in the MIPSGAL image. 

\subsection{Distance}

We are studying a region in the first Galactic quadrant, thus we have to take into account the distance 
ambiguity that exists when using radial velocities and a Galactic rotation curve to assign distances 
to sources.
According to \citet{kolpak03} and \citet{anderson09}, G45.45+0.06 has a radio recombination line velocity of 
$\sim$ 56 \k, which gives the possible distances of $\sim 4$ or $\sim 8$ kpc. As the HI absorption spectrum 
towards G45.45+0.06 presents an absorption feature at the tangent point velocity ($\sim 64$ \k), 
the authors adopted the farther distance for G45.45+0.06. 

\begin{figure}[h]
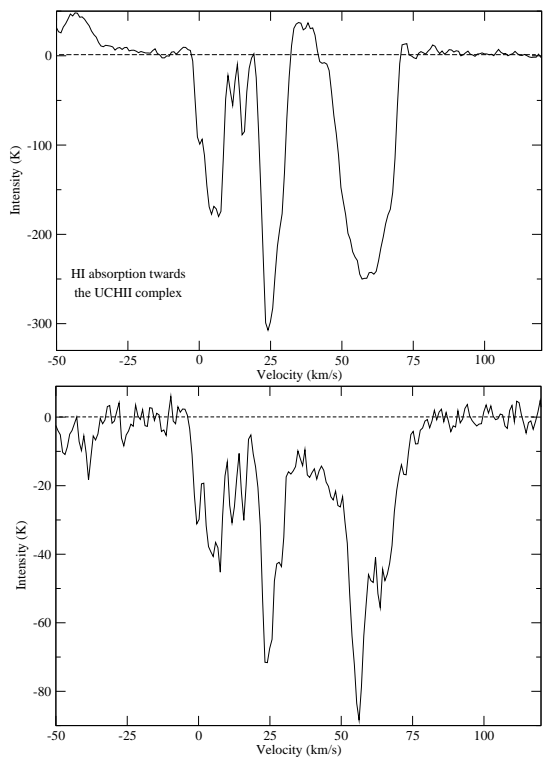

\centering
\includegraphics[totalheight=0.2\textheight,viewport=0 0 700 470,clip]{absUCHIIcomplex.eps}
\includegraphics[totalheight=0.2\textheight,viewport=-14 0 700 470,clip]{absNewHII.eps}
\caption{Up: HI absorption towards the UCHII complex where G45.45+0.06 is embedded. Down: HI absorption 
towards the radio continuum structure enclosed by North and middle PDR. Note that both profiles have the 
same HI absorption features.}
\label{abs}
\end{figure}

The molecular gas belonging to the shell associated with G45L (see Figure \ref{present}) is at the same velocity 
range as the gas related to the UCHII complex where G45.45+0.06 is embedded, which may suggest
that G45L is at the same distance as the UCHII complex. In order to prove this suggestion, 
using HI data extracted from the VLA Galactic Plane Survey (VGPS) \citep{stil06}, we studied 
the absorption features towards two regions: the UCHII complex where G45.45+0.06 is embedded (Figure \ref{abs}-up) 
and the radio continuum structure enclosed by North and middle PDR (Figure \ref{abs}-down). 
As both profiles have the same HI absorption features, we conclude that G45L is at the 
same distance as G45.45+0.06. Hereafter we adopt the distance of 8 kpc.

\subsection{Molecular Environment}
\label{molec}

\begin{figure}[h]
\centering
\includegraphics[width=9.5cm]{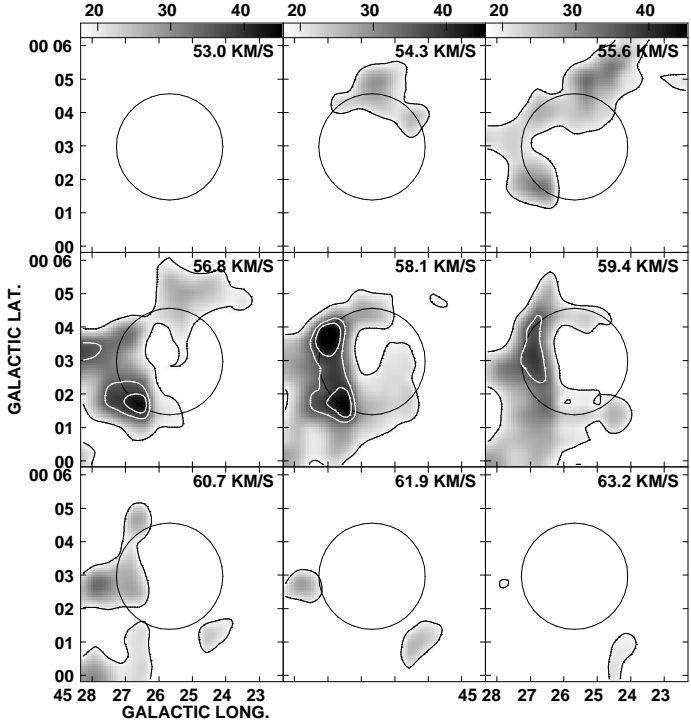}
\caption{Integrated velocity channel maps of the \3 J=1--0 emission every $\sim$ 1.25 \k. 
The grayscale is displayed at the top of the figure and is in K \k, the contour levels are 5, 9.5 and 11 K \k.
Note that the molecular gas encompasses the circle, which highlights the position and size of G45L as suggested 
by the North and South PDRs (see Figure \ref{combi}-left). }
\label{panel}
\end{figure}

As shown in Figure \ref{present}, an incomplete \3 shell at the velocity range between 50 and 65 \ks presents a good 
morphological correlation with the North and South PDRs visible in the 8 $\mu$m band (see Figure \ref{combi}-left). 
This suggests that the expansion of the HII region, that we called G45L, is collecting the molecular gas as is 
observed in other HII regions (e.g. \citealt{poma09,silvina09,deha08}). Figure \ref{panel} shows the integrated velocity 
channel maps of the \3 J=1--0 emission every $\sim$ 1.25 \k. For reference, the circle highlights the position and 
the size of G45L as suggested by the North and South PDRs. From this figure is 
clear that the molecular gas is more abundant towards the East. In particular, the v $= 58.1$ \ks channel map 
shows two molecular clumps spatially coincident with the UCHII complex and the region where 
the EGO G45.47+0.05 lies, respectively. It is also evident that 
the molecular gas encompasses the circle, showing the velocity structure of the molecular shell related to G45L.
As \citet{deha05} propose, the presence of a molecular shell surrounding the ionized gas of an HII region, or the 
presence of massive fragments regularly spaced along the ionization front, suggest that the 
collect and collapse process is at work in the region.
On the other hand, the channels with velocities of 56.8 and 58.1 \ks show a smaller molecular shell 
interior to the circle and in spatial coincidence with the structure delimited by the middle and North PDRs 
(see Figure \ref{combi}-left). 

\begin{figure}[h]
\centering
\includegraphics[width=8cm]{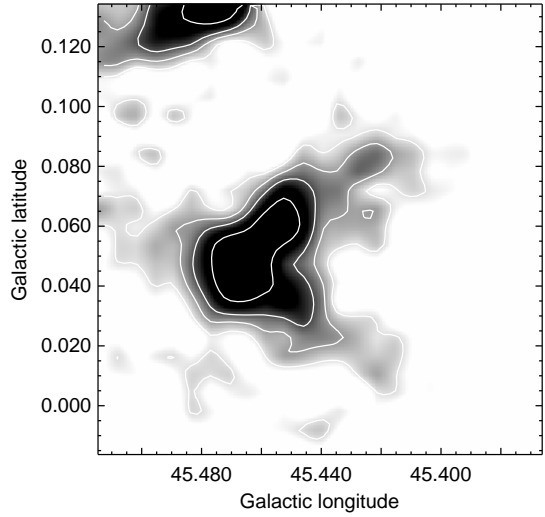}
\caption{CS J=2--1 emission integrated between 54 and 66 \k. The contour levels are 2.4, 3.6, 5 and 8 K \k. 
Note that the CS emission presents a similar structure as the \3.}
\label{csinteg}
\end{figure}

Figure \ref{csinteg} displays the CS J=2--1 emission integrated between 54 and 66 \k. 
The CS J=2--1 emission presents a similar structure 
as the \3. Given that this line requires high densities, $10^4 - 10^5$ cm$^{-3}$, to be excited 
(see e.g. \citealt{luna06}), its detection indicates the presence of high density gas in the molecular shell, 
mainly towards the East. 

\begin{figure*}[h]
\centering
\includegraphics[width=14.2cm]{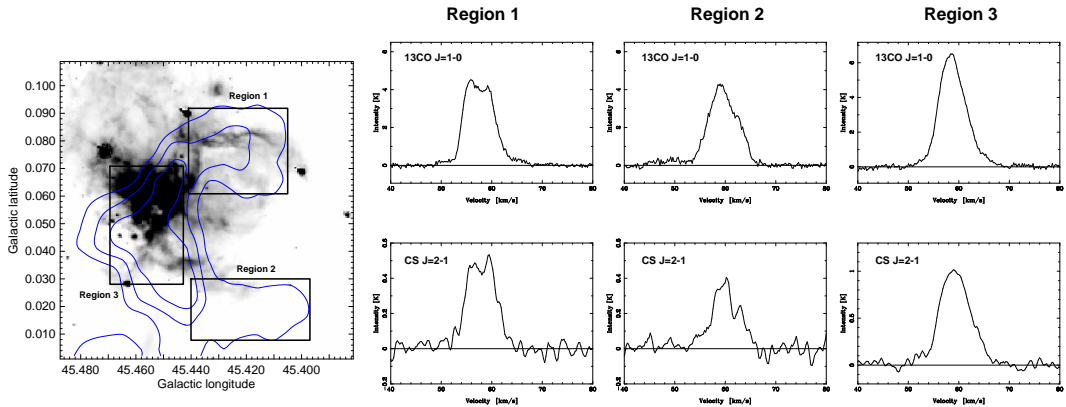}
\caption{Left: {\it Spitzer}-IRAC 8 $\mu$m emission with contours of the integrated
\3 J=1--0 as presented in Figure \ref{present}. The boxes represent the regions from which were obtained the 
average of the molecular emissions. Towards the right the  
\3 J=1--0 and CS J=2--1 average spectra from each region are shown. The rms noise are $\sim 0.03$ 
and $\sim 0.04$ K for the \3 and CS, respectively.}
\label{espectrosAve}
\end{figure*}

In order to analyze the kinematics of the molecular gas,
we study the molecular spectra from three different regions: Region 1 (the North portion of the molecular shell, 
in coincidence with the North PDR), 
Region 2 (the South portion of the molecular shell, in coincidence with the South PDR) 
and Region 3 (where the molecular emission peaks). 
Figure \ref{espectrosAve} (left) shows the {\it Spitzer}-IRAC 8 $\mu$m emission with contours of the integrated 
\3 J=1--0 as presented in Figure \ref{present}. Regions 1, 2 and 3 are indicated with boxes of approximately 
2\m~$\times$ 1\farcm8, 2\farcm2 $\times$ 1\farcm2 and  1\farcm5 $\times$  2\farcm5 in size, respectively.  
Towards the right, 
the \3 J=1--0 and CS J=2--1 average spectra corresponding to the emission of each region are displayed. Between 
40 and 50 spectra were averaged to obtain each average emission spectrum.
The parameters determined from Gaussian fitting of these lines are presented in Table \ref{avelines}.
T$_{mb}$ represents the peak brightness temperature, V$_{lsr}$ the central velocity,
$\Delta v$ the line width and $I$ the integrated line intensity. Errors are formal 1$\sigma$ value for the model
of the Gaussian line shape. The \3 and CS average lines corresponding to Region 1 were best fitted with two Gaussians. 

\begin{table}[h]
\caption{Observed parameters of the \3 J=1--0 and CS J=2--1 average emissions towards the regions shown 
in Figure \ref{espectrosAve}.}
\label{avelines}
\centering
\begin{tabular}{ccccc}
\hline\hline
Emission & T$_{mb}$ & V$_{lsr}$ & $\Delta v$  & $ I $    \\
         &  (K)     & (\k)      &   (\k)      & (K \k)                \\
\hline
\multicolumn{5}{c}{\it Region 1} \\          
\3 J=1--0 & 3.80 $\pm$0.50  &  58.75 $\pm$1.05 & 5.05 $\pm$0.25 & 17.00 $\pm$3.00 \\
          & 4.10 $\pm$0.50  &  55.25 $\pm$1.05 & 3.00 $\pm$0.15 & 15.00 $\pm$2.00 \\
CS J=2--1 & 0.46 $\pm$0.11  &  59.60 $\pm$1.10 & 4.00 $\pm$0.20 & 2.00 $\pm$0.30 \\
          & 0.42 $\pm$0.15  &  55.80 $\pm$1.10 & 3.45 $\pm$0.15 & 1.90 $\pm$0.30 \\
\hline

\multicolumn{5}{c}{\it Region 2} \\
\3 J=1--0 & 4.10 $\pm$0.50  &  59.50 $\pm$0.50 & 6.75 $\pm$0.85 & 30.00 $\pm$3.80 \\
CS J=2--1 & 0.35 $\pm$0.05  &  59.90 $\pm$0.50  & 6.25 $\pm$0.95 & 2.30 $\pm$0.50 \\
\hline
\multicolumn{5}{c}{\it Region 3} \\
\3 J=1--0 & 6.40 $\pm$0.50  &  58.65 $\pm$0.25 & 6.10 $\pm$0.50 & 41.50 $\pm$3.50 \\
CS J=2--1 & 1.00 $\pm$0.10  &  59.25 $\pm$0.55  & 7.00 $\pm$1.00 & 7.70 $\pm$1.50 \\
\hline
\end{tabular}
\end{table}

\begin{figure}[h]
\centering
\includegraphics[width=4.2cm,angle=-90]{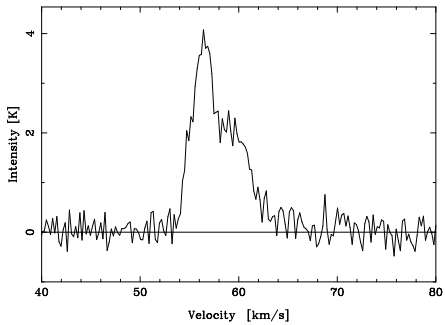}
\caption{\3 J=1--0 spectrum obtained towards the middle PDR, at $l = $45\fdg$430, b = $0\fdg$055$. The 
main component velocity is v $\sim 56$ \k, which coincides with one of the two molecular components of the North PDR.} 
\label{13PDRmid}
\end{figure}

Besides, taking into account that the \3 J=1--0 beam (46\s) covers completely the middle PDR, 
we present in Figure \ref{13PDRmid} 
a spectrum obtained from  $l = $45\fdg$430, b = $0\fdg$055$. This spectrum shows the molecular 
emission probably related to the middle PDR. The velocity of the main component of 
this profile is v $\sim 56$ \k, in coincidence with one of the two molecular components of the North PDR 
(Region 1 in Figure 
\ref{espectrosAve}). The spectrum is not symmetrical: it presents another weaker
component or a wing towards larger velocities.

The molecular spectra analysis shows that the North PDR has associated two possible molecular structures centered 
at v $\sim 55$ \ks and v $\sim 59$ \k, respectively. The main velocity component of the molecular 
gas associated with the middle PDR is centered at v $\sim 55$ \k, while the molecular gas related to the South 
PDR is centered at v $\sim 59$ \k.
If the molecular gas is indeed associated with G45L, this may suggest that we have to take into account 
projection effects to understand the three dimensional structure of G45L. This will be discussed in 
Section \ref{estruct}.

Using the \3 J=1--0 line and assuming local thermodynamic equilibrium (LTE) we estimate the H$_{2}$ column density
towards the three regions shown in Figure \ref{espectrosAve}. We use:
$${\rm N(^{13}CO)} = 2.42 \times 10^{14} \frac{T_{\rm ex} \int{\tau_{13} dv}}{1 - exp(-5.29/T_{\rm ex})}$$
to obtain the \3 column density. $\tau_{13}$ is the optical depth of the line and 
following \citet{anderson09} the  $T_{\rm ex}$ was assumed to be 13 K. 
We assume the \3 emission is optically thin and use the Gaussian fit line parameters (Table \ref{avelines})
to find the optical depth integral. Finally, we use the relation N(H$_{2}$)/N(\3)$ \sim 5 \times 10^5$ 
(e.g. \citealt{simon01}) to estimate the following values: 
N(H$_{2}$) $\sim 1.5 \times 10^{22}$ cm$^{-2}$ for Region 1 and 2, respectively and 
N(H$_{2}$) $\sim 2.8 \times 10^{22}$ cm$^{-2}$ for Region 3.

In addition we estimate the CS column densities, N(CS), towards these regions.
Although the CS J=2--1 line is generally optically
thick, a lower limit to the N(CS) can be estimated under the assumption of optically thin CS emission. We use 
the following equation \citep{ohashi91}:
$$ {\rm N(CS)} = 8.5 \times 10^{11} \frac{exp(2.4/T_{\rm ex})}{1 - exp(-4.7/T_{\rm ex})} 
\times {T_{\rm mb}} \times \Delta v, $$ where
$T_{\rm mb}$ is the CS brightness temperature and $\Delta v$ is the velocity width of the CS line.
Following \citet{goico06}, who studied the Horsehead PDR using CS J=2--1 between other molecular lines, 
we assume T$_{\rm ex} = 9$ K. We obtain N(CS) $\sim$ (8 and 6) $\times 10^{12}$ cm$^{-2}$ for Region 1 and 2, 
respectively, and  N(CS) $\sim 2\times 10^{13}$ cm$^{-2}$ for Region 3.

From the H$_{2}$ and CS column densities, we can estimate the CS abundances, $X({\rm CS}) = {\rm N(CS)/N(H_{2})}$.
We obtain $X({\rm CS}) \sim$ (5, 3.5 and 7) $\times 10^{-10}$ for Region 1, 2 and 3, respectively.
Taking into account the LTE approximation and that the estimated CS column densities are lower limits, these values
are comparable with those obtained towards the Orion Bar (a warm PDR), $X({\rm CS}) \sim 2.9 \times 10^{-9}$ 
\citep{johnstone03} and towards the Horsehead PDR, $X({\rm CS}) \sim (7 \pm 3) \times 10^{-9}$ \citep{goico06}.
New CS and C$^{34}$S observations in several lines would be very useful to study the sulfur depletion in G45L PDRs.

Finally, we estimate the total mass of the whole molecular shell in M $\sim$ 10$^{5}$ \msol. This value
was obtained from:
$$ {\rm M} = \mu~m_{{\rm H}} \sum{\left[ D^{2}~\Omega~{\rm N(H_{2})} \right] }, $$
where $\Omega$ is the solid angle subtended by the \3 J=1--0 beam size, $m_{\rm H}$ is the hydrogen mass, 
$\mu$, the mean molecular weight, is assumed to be 2.8 by taking into account a relative helium abundance 
of 25 \%, and $D$ is the distance. Summation was performed over all the observed positions within the 
27 K \ks contour level (see Figures \ref{present} and \ref{espectrosAve}).

\subsection{The exciting star(s) of G45L}

\begin{figure}[h]
\centering
\includegraphics[width=9cm]{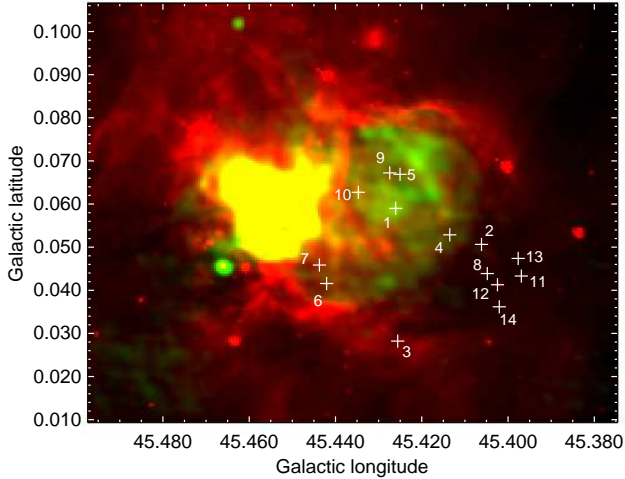}
\caption{Colour composite image, with the radio continuum emission at 1420 MHz in green, and the infrared
emission at 8 $\mu$m in red. Yellow is the superposition of the IR
and radio continuum emissions. The exciting stars candidates are identified (see Table \ref{irsourcesT}).}
\label{ir1}
\end{figure}

\begin{figure}[h]
\centering
\includegraphics[width=9cm,angle=-90]{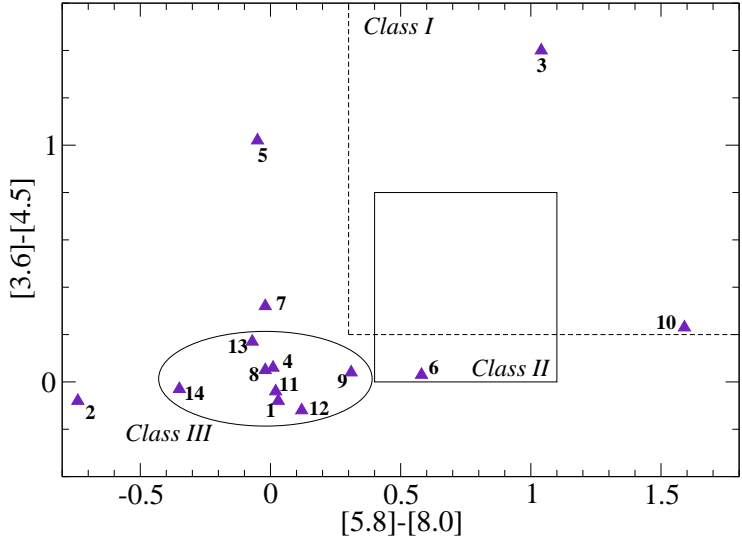}
\caption{GLIMPSE-IRAC color-color diagram [3.6] -- [4.5] versus [5.8] -- [8.0] for the sources displayed 
in Figure \ref{ir1}. Class I, II and III regions are indicated following \citet{all04}.
The ellipse (Class III region: main sequence and giant stars) encloses the sources that we consider to be possible 
exciting star(s) of G45L.} 
\label{ir2}
\end{figure}

Given that we do not find any cataloged massive star in the area that may be related to G45L, we have used
the GLIMPSE I Spring'07 Catalog to analyze the infrared sources seen in projection onto G45L
to identify the star(s) responsible for its creation.
Considering only the sources that have been detected in the four Spitzer-IRAC
bands, we found 14 sources towards the region located inside the observed PDRs.
The main parameters of these sources are shown in Table \ref{irsourcesT}.
Figure \ref{ir1} indicates the location of the sources with respect to the emission distribution at 8 $\mu$m (red)
and the radio continuum at 20 cm (green).
To examine the evolutionary stage of these sources, we analyze their location onto a color-color IRAC diagram as
shown in Figure \ref{ir2}.
Following \citet{all04} color criteria, we found that 8 sources could be classified as main
sequence stars (Class III). Among these sources we look for O-type stars as responsible for ionizing the surrounding gas. 
Using the $J$, $H$ and $K$ apparent magnitudes as obtained from the 2MASS Point Source Catalog, 
we estimate the absolute magnitudes for the mentioned 8 sources. To convert the apparent
magnitudes in absolute ones, we assume a distance of 8 kpc and a visual absorption between 8 and 12 mag.
The extinction values were obtained by inspecting the infrared sources location 
onto the color-color diagram {\it (H-K)} vs {\it (J-H)} (not presented here).
By comparing the estimated absolute magnitudes with those tabulated by \citet{mar06},
we found that sources \#1, \#9 and \#14 are probably O-type stars (likely between O4V and O8V), 
which is consistent  with their position in the color-magnitude diagram 
{\it K} vs {\it (H-K)} (not presented here). Source \#1 is slightly displaced towards the region 
of giants stars in the color-magnitude diagram, but it could be a reddened O-type star.  
An inspection of Figure \ref{ir1} shows that sources \#1
and \#9 are located almost onto the central part of G45L, while source \#14 lies towards its boundary.
This fact suggest that sources \#1 and \#9 are the most promising candidates for being related to G45L.
In particular, source \#9 is projected onto a local minimum observed in the brightest part of the G45L radio continuum
emission, suggesting that this source could have been blowing its environs creating a small cavity around it.

\begin{table*}
\tiny
\caption{Main parameters of the infrared sources found towards G45L.}
\label{irsourcesT}
\begin{tabular}{lcccccccccc}
\hline
\hline
\# & Glimpse designation & Galactic coordinates & $J$ & $H$ & $K{_S}$ & 3.6 $\mu$m & 4.5 $\mu$m &  5.8 $\mu$m & 8.0 $\mu$m & Notes\\
\hline
1 & G045.4260+00.0590	& 45\fdg426, 0\fdg058	& 13.93	& 11.89	& 11.08 &  10.47 &  10.55 &  10.21 &  10.18 & Class III - (O-type)\\
2 & G045.4061+00.0506	& 45\fdg406, 0\fdg050	& 13.13	& 11.67	& 11.08	& 10.71	&  10.79 &  10.58 &  11.32 & \\
3 & G045.4255+00.0282	& 45\fdg425, 0\fdg028	& null	& null	& null	& 13.7	&  12.3	&  11.09 &  10.05 & \\
4 & G045.4135+00.0528   & 45\fdg413, 0\fdg052	& 11.51	& 10.45	& 10.09	& 9.86	&  9.8	&  9.69	&  9.68 & Class III\\
5 & G045.4250+00.0668   & 45\fdg425, 0\fdg066	& null	& null	& 14.43	& 11.33	&  10.31 &  9.77 &  9.81 & \\
6 & G045.4420+00.0415   & 45\fdg442, 0\fdg041	& null	& null	& null	& 12.07	&  12.03 &  11.64 &  11.06 & \\
7 & G045.4437+00.0458	& 45\fdg443, 0\fdg045	& null	& null & 12.67	& 10.68	&  10.36 &  9.81 &  9.84 & \\
8 & G045.4048+00.0438	& 45\fdg404, 0\fdg043	& null	& 13.65 & 12.17	& 11.04	&  10.99 &  10.58 &  10.6 & Class III\\
9 & G045.4274+00.0672	& 45\fdg427, 0\fdg067	& 13.11	& 11.64 & 11.07	& 10.53	&  10.49 &  9.94 &  9.63 & Class III - (O-type)\\
10 & G045.4347+00.0627	& 45\fdg434, 0\fdg062	& null	& null & null	& 12.77	&  12.54 &  9.40 &  7.81 & \\
11 & G045.3969+00.0433   & 45\fdg396, 0\fdg043	& 11.79	& 10.41	& 9.85	& 9.55	&  9.6	&  9.41	&  9.39 & Class III\\
12 & G045.4024+00.0412   & 45\fdg402, 0\fdg041	& 10.41	& 9.21	& 8.73	& 8.39	&  8.51	&  8.34	&  8.22 & Class III\\
13 & G045.3976+00.0473   & 45\fdg397, 0\fdg047	& null	& 13.67	& 12.42	& 11.5	&  11.33 &  10.82 &  10.89 & Class III\\
14 & G045.4020+00.0361  & 45\fdg402, 0\fdg036	& 13.42	& 12.28	& 11.88 & 11.52	&  11.55 &  11.38 &  11.73 & Class III - (O-type)\\
\hline
\end{tabular}
\end{table*}

In order to investigate if these three sources can provide the energy necessary to ionize the gas, we need to
estimate the radio continuum flux of G45L.
Using the radio continuum emission data at 1420 MHz, we estimate a flux density of
$S_{1420 MHz} \sim 1.0 \pm 0.2$ Jy for G45L.
The number of UV photons necessary to keep the gas ionized is derived using
$N_{UV}(\rm photons\, s^{-1}) = 0.76 \times 10^{47}\, T_4^{-0.45}\,
\nu_{\rm GHz}^{0.1}\, S_{\nu}\, D_{\rm kpc}^2$ \citep{cha76}, where $T_4$ is the electron temperature in units of
10$^4$ K, D$_{\rm kpc}$ is the distance in kpc
and  $S_{\nu}$ is the flux density in Jy. Adopting an electron temperature of 10$^4$ K, we obtain
$N_{UV} = (5 \pm 2) \times 10^{48} \rm s^{-1}$. On the other hand,
the UV photon flux corresponding to the contribution of the three stars is $\sim 3.3 \times 10^{49}$
photons per second  \citep{mar05}.
Thus, we conclude that the three stars can maintain the HII region G45L ionized and heat the dust emitting at
24 $\mu$m in the ionized gas.
It is important to mention that one of the central sources by itself, \#1 or  \#9,
can provide the necessary UV photons. On the contrary, source \#14 can not have created the HII region G45L alone.

Assuming that sources \#1, \#9, and \#14 are the sources responsible for the HII region G45L,
we estimate its dynamical age
using a simple model described by \citet{dys80}.
In this model the radius of the HII region at a given time $t$ is given by
$ R(t) = R_s\,(1 + 7 \, c_s\, t /\, 4 \, R_s)^{4/7}$, where $c_s$ is the sound velocity in the ionized gas
($c_s = 15$ \k) and $R_s$ is the radius of the
Str\"omgren sphere, given by $R_s = (3\, S_* /4\, \pi\, n_o^2\, \alpha_B)^{1/3}$, where
$\alpha_B = 2.6 \times 10^{-13}$ cm$^3$ s$^{-1}$ is the hydrogen recombination coefficient to all levels above
the ground level. $S_*$  is the total number of ionizing photons per unit of time emitted by the stars,
and $n_o$ is the original ambient density.
As a rough estimate $n_o$
can be obtained by distributing the total molecular mass related to the structure (M $\sim 10^{5}$ \msol)
over a sphere of about 7 pc
(3\m~at 8 kpc) in diameter, which yields $n_{0} \sim 10^{4}$ cm$^{-3}$.
Given that the actual diameter of G45L is about 7 pc, we infer that the HII region has been
expanding during about $2 \times 10^6$ yr.

\subsection{Star formation around G45L}

\begin{figure}[h]
\centering
\includegraphics[width=8cm]{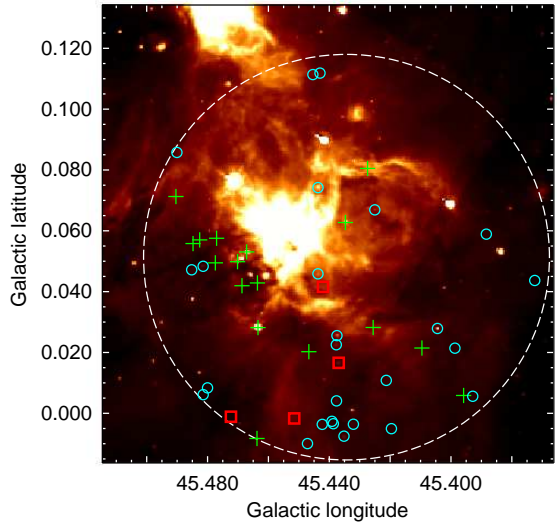}
\caption{{\it Spitzer}-IRAC 8 $\mu$m emission with YSO candidates superimposed. Green crosses indicate Class I sources,
red boxes are Class II or intermediate Class I/II sources and the cyan circles are the sources that could be
reddened Class II objects. The circle encloses the region where the photometric study was performed.}
\label{ysos}
\end{figure}

\begin{figure}[h]
\centering
\includegraphics[width=8cm,angle=-90]{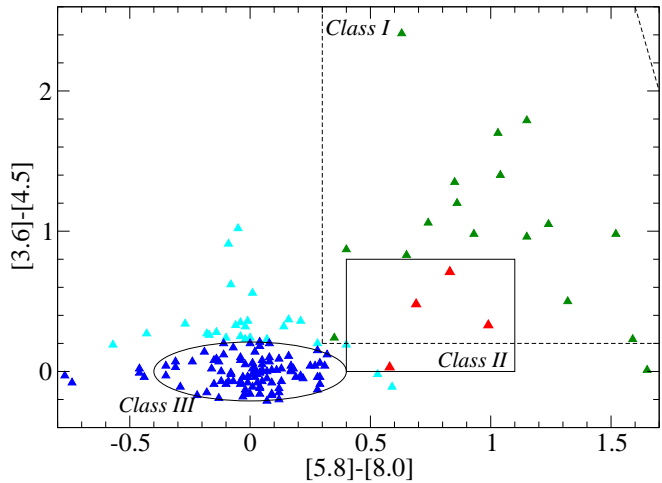}
\caption{GLIMPSE-IRAC color-color diagram [3.6] -- [4.5] versus [5.8] -- [8.0] for the sources observed inside
the dashed circle displayed in Figure \ref{ysos}. As in Figure \ref{ir2}, Class I, II and III regions are 
indicated following \citet{all04}. In this case we consider Class I (green), Class II and intermediate 
Class I/II (red) and reddened Class II (cyan) objects to study star formation around G45L.} 
\label{ccysos}
\end{figure}

To look for primary tracers of star formation activity around G45L,
we use the GLIMPSE I Spring'07 Catalog to perform photometry.
Considering only sources that have been detected in the four {\it Spitzer}-IRAC bands, we found 151 sources in
the area delimited by the dashed circle shown in Figure \ref{ysos}. The circular area was chosen in order 
to cover the molecular gas that surrounds the HII region and where the YSO candidates are expected to be located.
This figure displays the spatial distribution of the YSO candidates over the {\it Spitzer}-IRAC 8 $\mu$m 
emission. The green crosses indicate Class I sources, the
red boxes indicate Class II or intermediate Class I/II sources and the cyan circles indicate the sources that
could be reddened Class II objects. The sources classification was performed according the photometric study presented 
in Figure \ref{ccysos}, which shows the IRAC color-color diagram of the sources found.
The different regions indicated correspond to different stellar evolutionary stages, as defined 
by \citet{all04}.
Seventeen sources lie in the region of the protostars with circumstellar envelopes
(Class I, green triangles),
only 4 sources lie in the region of
young stars with only disk emission (Class II and intermediate Class I/II, red triangles) and 104 sources
lie in the region of the main sequence and giant stars (Class III, blue triangles).
Sources represented as cyan triangles, located outside the
delimited regions, could therefore be reddened Class II objects \citep{all04}.
Unfortunately we can not perform
an additional NIR photometric study because the 2MASS data from most of these sources are either 
missing or given as lower limits.

\begin{figure*}[h]
\centering
\includegraphics[width=15cm]{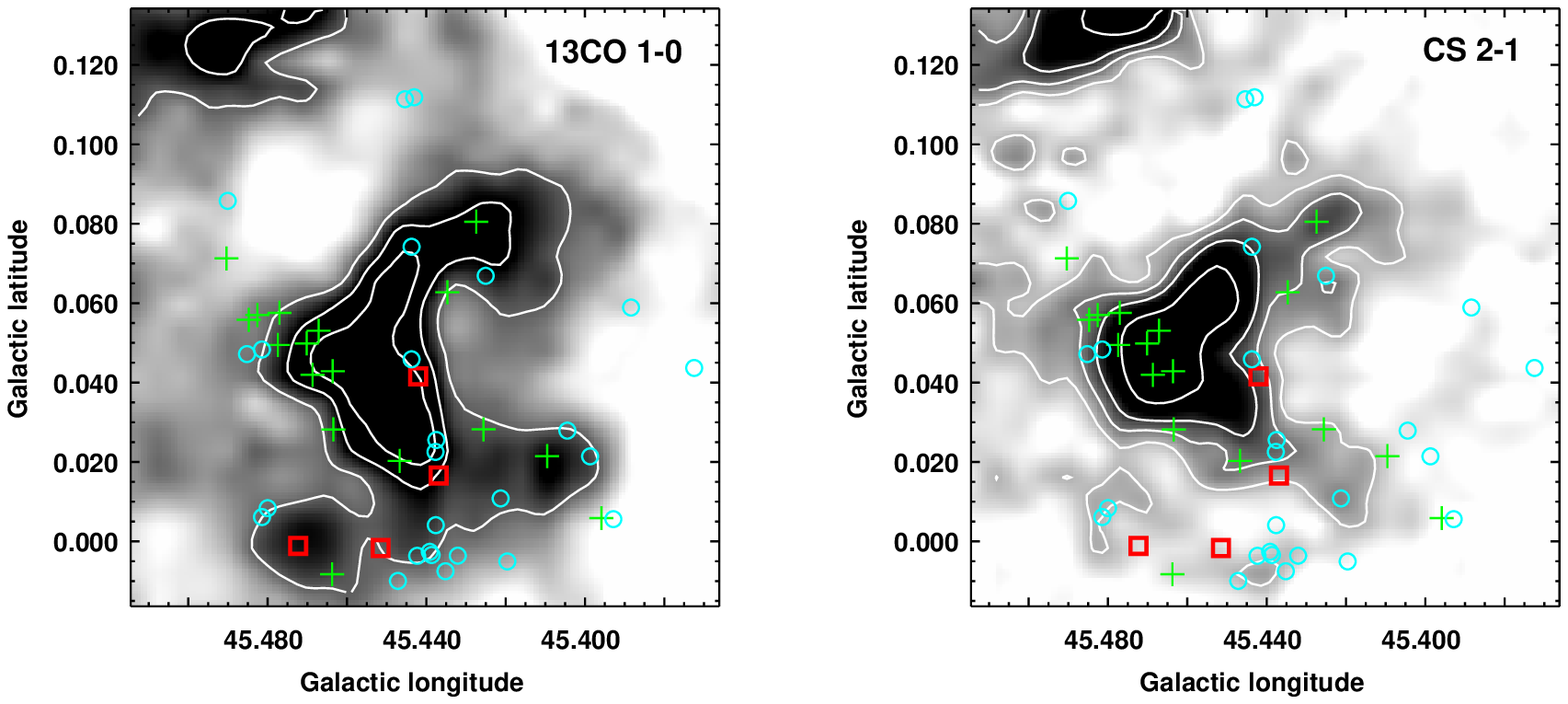}
\caption{Left: \3 J=1--0 integrated emission between 54 and 66 \ks with YSO candidates superimposed. The contour 
levels are 27, 37 and 48 K \k. Right: same, but grays are the CS J=2--1 integrated emission in the same velocity
range. The contour levels are 2.4, 3.6, 5 and 8 K \k. Green crosses indicate Class I sources,
red boxes are Class II or intermediate Class I/II sources and the cyan circles are the sources that could be 
reddened Class II objects.}
\label{ysos+molec}
\end{figure*}

In Figure \ref{ysos+molec} the same sources shown in Figure \ref{ysos} are displayed over
the \3 J=1--0 (left) and CS J=2--1 (right) integrated emissions between 54 and 66 \k.
Of course we do not know if all the sources seen towards G45L are at the same distance as the
HII region. However, the location of some of them, mainly the Class I objects (green crosses in Figure
\ref{ysos+molec}), suggest that they are possibly embedded
in the molecular gas related to G45L. The Class I objects lie preferently in the molecular shell and concentrate
towards the East, where the CS J=2--1 emission peaks. They lie in the region
where the EGO G45.47+0.05 is embedded \citep{cyga08}, confirming that in this area, southeastern the UCHII complex and
eastern G45L, star formation is taking place.

As mentioned in Section 1, \citet{feldt98} and \citet{blum08} identified several massive O-type stars,
that are presumably on, or near, the zero-age main sequence, and are
responsible for the ionization of the UCHII region G45.45+0.06. The authors also proposed that this
UCHII region triggered the formation of younger UCHII regions in its surroundings, generating
an UCHII complex. These studies do not mention any agent responsible for the formation of such stars.
Taking into account the estimated age of G45L (about $2 \times 10^6$ yr),
that star formation is taking place around it, probably through the collect and collapse process, and
that the UCHII complex lies on one of its border, we propose that G45L could have triggered the
formation of the zero-age main sequence stars that are ionizing G45.45+0.06. However we can not discard
the possibility that both HII regions, G45.45+0.06 and G45L, could be coeval.

\subsection{HII region G45L spatial structure}
\label{estruct}

In this section, based on the IR emission and the molecular environment study, we attempt to describe the possible 
spatial structure of G45L. 

\begin{figure}[h]
\centering
\includegraphics[width=6cm]{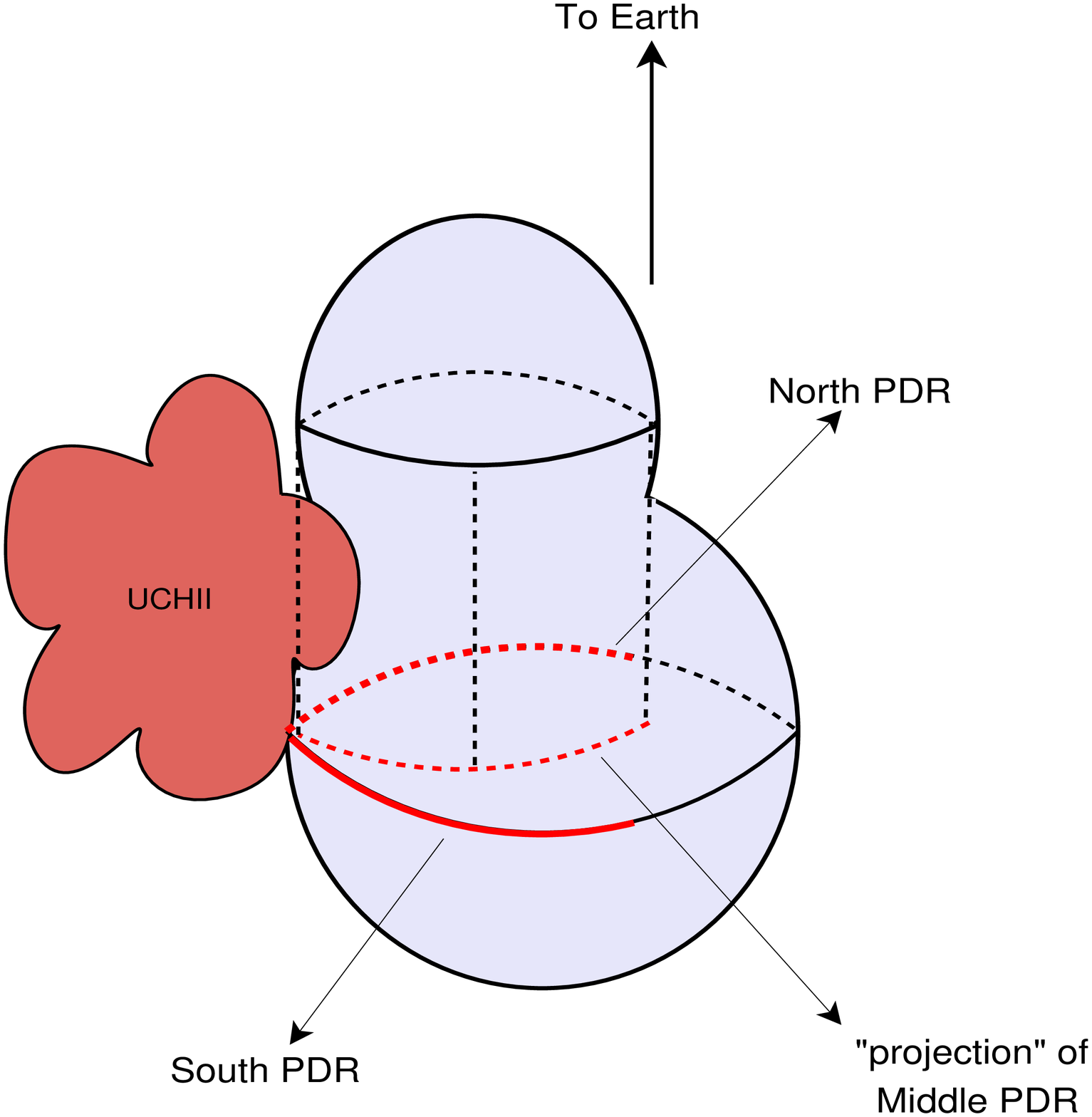}
\includegraphics[width=6cm]{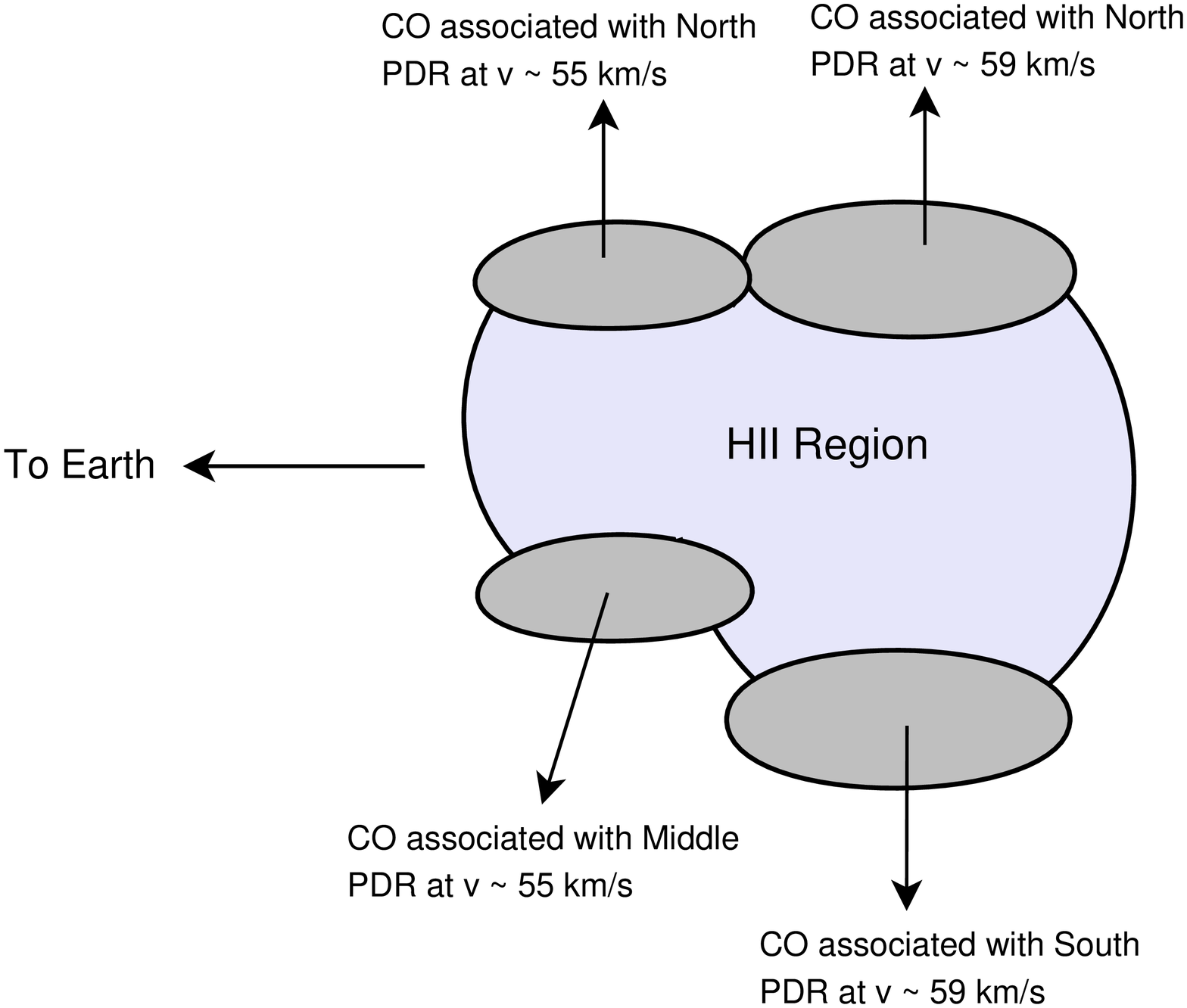}
\caption{Left: sketch of the possible shape of the HII region G45L. The PDRs are remarked. The UCHII complex is
represented as a red cloud. Right: a longitudinal cut of G45L remarking the molecular gas components related to
the PDRs.}
\label{esquema}
\end{figure}

As described in previous sections, a view of a larger area in the vicinity of the UCHII region G45.45+0.06 reveals 
that it actually lies in a border of a larger and fainter HII region that we called G45L.
The morphological study of G45L suggests that this HII region is far from being spherical.
The observed morphology is quite complex and then projection effects should be considered to 
analyze the three dimensional structure.
In this way, taking into account the location of the PDRs together with the spatial and velocity molecular 
gas distribution, we suggest that G45L has a pear-like morphology, as it is sketched in Figure \ref{esquema}.
This figure presents two sketches of G45L possible structure as seen from two different positions.
In Figure \ref{esquema} (left) can be appreciated that the North PDR is composed 
by two structures seen superimposed along the line of sight and the middle PDR is the projection of the southern border
of the HII region portion closer to us. The UCHII complex that lies towards the East of G45L is represented as 
a red cloud. Figure \ref{esquema} (right) shows a longitudinal cut of G45L remarking the molecular gas components 
related to the PDRs.
As the molecular analysis shows, the structure delimited by the North and the middle PDRs has
associated molecular gas centered at v $\sim 55$ \k, while the structure delimited by the South and
the North PDRs has associated molecular gas centered at v $\sim 59$ \k.

Most ionized nebulae, in particular HII regions, have complex morphologies structures \citep{morisset05}.
There are many physical causes that may account for such morphologies. Some causes 
for the G45L spatial structure could be: the presence of density gradients in the 
ISM where the HII region is evolving, the existence of more than one exciting star, the fact that these 
exciting stars could have high spatial velocities with respect to the local ISM, and the 
effects of possible stellar and/or interstellar magnetic fields.

\section{Summary}

Using multiwavelength survey and archival data, we studied the ISM towards a region 
about 7\m $\times$ 7\m~in the vicinity of the G45.45+0.06 UCHII complex. The main results can be summarized 
as follows:

(a) We found that the UCHII complex lies in a border of a larger ($\sim$ 3\m~of diameter) and 
fainter HII region, here named G45L.

(b) Although G45L is not completely border by a PDR, its morphology resembles the structure of
the IR dust bubbles associated with O and early-B type stars: a PDR visible in
the 8 $\mu$m band, which encloses ionized gas observed at 20 cm and hot dust observed at 24 $\mu$m.

(c) We find a good morphological correlation between the PDRs 
and the molecular gas, which suggests that the HII region may be collecting the molecular material.

(d) Taking into account the velocity (v$_{\rm LSR} \sim 55 - 60$ \k) of the molecular gas related to G45L
and the HI absorption study, we conclude that this HII region is at the same distance as the UCHII 
complex, $\sim 8$ kpc.

(e) The PDRs position and the spatial and velocity distribution of the associated molecular shell 
suggest that G45L has a pear-like morphology.

(f) From a near- and mid-IR photometric study, we found three sources, likely O-type stars (between O4V and O8V)
that are possibly responsible for the creation of G45L. 
Additionally we found several YSO candidates lying preferently in 
the molecular shell and concentrating towards the East where the CS J=2--1 emission peaks. 
Our results confirm that the southeastern region 
of the UCHII complex where G45.45+0.06 is embedded (eastern part of G45L) is an active star formation region. 

(g) Assuming that three O-type stars are responsible of G45L, 
we suggest this HII region has been expanding during 
about $2 \times 10^6$ yr and could have triggered the formation of the zero-age main sequence stars 
that are ionizing the UCHII region G45.45+0.06.
However we can not discard that both HII regions are coeval.

\begin{acknowledgements}
We wish to thank the anonymous referee whose comments and suggestions have helped to improve the paper.
S.P. and S.C. are members of the {\sl Carrera del 
investigador cient\'\i fico} of CONICET, Argentina. M.O. is a doctoral fellow of CONICET, Argentina. 
This work was partially supported by the CONICET grant PIP 112-200801-02166, UBACYT A023 and ANPCYT 
PICT-2007-00902 and -00812.

\end{acknowledgements}

%%%%%%%%%%%%%%%%%%%%%%%%%%%%%%%%%%%%%%%%%%%%%%%%%%%%%%%%%%%%%%%%%%%%%
\bibliographystyle{aa}  % A&A format
   %\bibliographystyle{klunamed}     
   % format of references provided by the review (.bst)
\bibliography{biblio}
   % file containing the bibtex references (.bib)
\IfFileExists{\jobname.bbl}{}
{\typeout{}
\typeout{****************************************************}
\typeout{****************************************************}
\typeout{** Please run "bibtex \jobname" to optain}
\typeout{** the bibliography and then re-run LaTeX}
\typeout{** twice to fix the references!}
\typeout{****************************************************}
\typeout{****************************************************}
\typeout{}
}

\end{document}